\documentclass[aps,preprint,dvipsnames]{revtex4-1}
\usepackage[french,english]{babel}
\usepackage{amsmath,amssymb}
\usepackage{graphicx}
\usepackage{verbatim}
\usepackage{color}
\usepackage{hyperref}
\usepackage{eqnarray,tabularx}
\usepackage{ccaption,float,esvect,amsthm,picins}
\usepackage{wrapfig}
\usepackage{amsbsy,amssymb,bm}
\usepackage{pstricks,pst-circ,pstricks-add,pst-node,pst-coil,pst-infixplot,pst-eucl,pst-text,pst-grad,pst-fun,pst-diffraction,pst-3d,pst-tools,pst-math,pst-solides3d}
\usepackage{amsmath}
\usepackage{subfigure,xcolor}

\begin{document}

\title{Curved wave rays in a ripple tank}
\author{\textcolor{blue}{Adel H. Alameh}}
\affiliation{\textcolor{blue}{Lebanese University, Department of Physics, Hadath, Beirut, Lebanon}}
\date{\today}
\email{adel.alameh@eastwoodcollege.com}

\begin{abstract}
In its broadest sense the term ``bent wave rays'' hints at light or electromagnetic waves bending in a strong gravitational field or in their progress through a transparent medium of nonuniform index of refraction. However, there are instances where curved wave rays are observed as a result of the superposition of two or more waves in an elastic medium.\\

 Amongst others, this paper aims at presenting a detailed examination of the phenomenon of interference of waves emanating from two synchronous and coherent sources and disturbing the surface of a liquid that is supposed ideally devoid of viscosity. Hence, it provides people with explicit parametric equations of the bent wave rays as well as those of their corresponding curved  wavefronts originating from the superposition of two waves in a material medium.\\

 To ensure the formality of the equations of the wave rays and  those of the corresponding wavefronts, a preliminary introduction to the mathematics of confocal conics is given in a quite intensive fashion.\\

  Therefrom, and for the sake of simplicity, and with no loss of generality, the study will be confined to transverse waves of the simple harmonic type having equal frequencies and equal amplitudes, thus, avoiding unnecessary complex situations that transcends the desired scope.
\end{abstract}
\maketitle

\subsection*{Confocal hyperbolae and ellipses}
 It is assumed that the reader is aware of the definition of an ellipse as being the set of points in a plane, the sum of whose distances from two fixed points $F_1$ and $F_2$ is equal to a constant. $F_1$ and $F_2$ are the foci of the ellipse. Literally,
 \begin{equation}r_1 +r_2=2a\label{literally}\end{equation}
   $r_1$ and $r_2$ being the radial distances separating a point of the ellipse from $F_1$ and $F_2$ respectively, and $a$ is the length of the semi major axis of the ellipse. In the cartesian system of coordinates the points of an ellipse satisfy the relation$\colon$
\begin{equation}\displaystyle\frac{x^2}{a^2}+\displaystyle\frac{y^2}{b^2}=1~\label{ellipse1}\end{equation}
 where $b$ is the length of the semi minor axis and whose value is determined by the relation,
 \begin{equation}b^2=a^2-c^2, \,\,\,\,\,\, \textrm{with}\,\, c<a ~\label{ellipse2}\end{equation}
   with $c$ representing the distance from the center $O$ of the ellipse to one of the foci.\\ Evidently, $F_1F_2=2c$\,.\\

  In a similar fashion, a hyperbola is defined to be the set of points in a plane, the difference of whose distances from two fixed points $F_1$ and $F_2$ is equal to a constant. $F_1$ and $F_2$ are also called the foci of the hyperbola. Hence the equation of a hyperbola in the bifocal form is given by the relation
  \begin{equation}r_2-r_1=2a\label{literal}\end{equation}
   $r_1$ and $r_2$ being the radial distances separating a point of the hyperbola from the two foci $F_1$ and $F_2$ respectively, and $2a$ is the distance separating the vertices of the two branches of the hyperbola. In the cartesian system of coordinates, the points of a hyperbola satisfy the relation$\colon$
  \begin{equation}\displaystyle\frac{x^2}{a^2}-\displaystyle\frac{y^2}{b^2}=1~\label{hyperbola1}\end{equation}
  with
  \begin{equation}b^2=c^2-a^2, \,\,\,\,\,\,  \textrm{with}\,\, c>a~\label{hyperbola2}\end{equation}
  Designating by $O$\,,  the midpoint of $F_1F_2$, implies that   $c=OF_1=OF_2$, and $F_1F_2=2c$.
  Now, a system of ellipses and hyperbolae having the same pair of foci are called confocal. Furthermore, ellipses and their corresponding confocal hyperbolae are certainly orthogonal. To put this fact into evidence we consider the following system of parametric equations of a point $M$ moving in a plane.
  \begin{equation}\left\{\begin{array}{lr}x=c \cosh u\cos v & ~\\
  y=c \sinh u\,\sin v & ~\\
  \end{array}\right.~\label{ellipsehyperbola}\end{equation}
  where $u$ and $v$ are respectively the angle made by the radial vector of tail at the origin and  head at the considered point $M$ in both cases of ellipses and  hyperbolae. See figure (1)\,.
\newpage

\hspace{1.5cm}\hfil\includegraphics[]{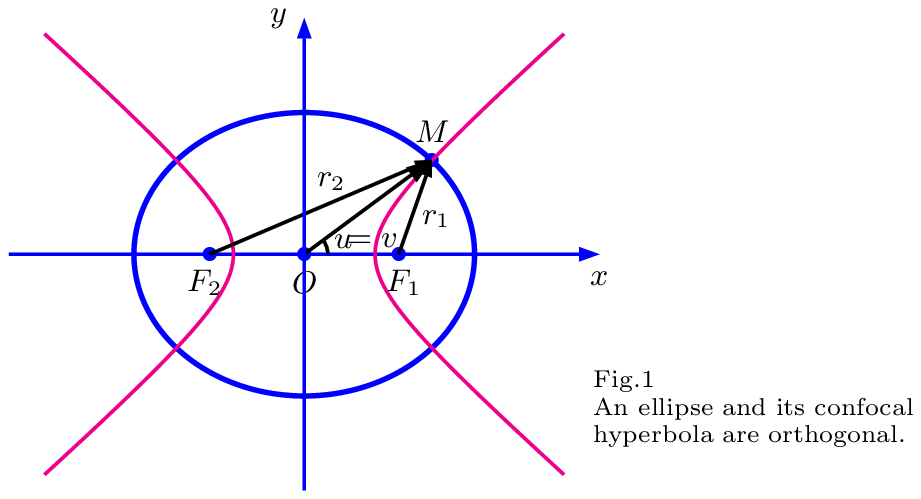}\hfil

\noindent If $u$ is seen as a parameter, and $v$ as a variable, then by defining the terms  $(a)$ and ($b$) for each numerical value $\eta$ of $u$, to be  the constants ($c\cosh \eta$)\,, and ($c\sinh \eta$) respectively,   equation~(\ref{ellipsehyperbola}) becomes$\colon$
\begin{equation}\left\{\begin{array}{lr}x=c\cosh \eta \cos v & ~\\
  y=c \sinh \eta \sin v & ~\\
  \end{array}\right.~\label{ellipse3}\end{equation}

 \noindent which is that of an ellipse, since by using $\cos^2 v+\sin^2 v=1$, and  $\cosh^2 u-\sinh^2 u=1$, we obtain equations~(\ref{ellipse1}) and~(\ref{ellipse2}) corresponding to an ellipse and its characteristic relation respectively. Likewise if $v$ is seen as a parameter, and $u$ as a variable, then by defining the terms ($a$) and ($b$) for each numerical value $\zeta$ of $v$  to be the constants ($c\cos \zeta$), and  ($c\sin \zeta$)\,, thus equation~(\ref{ellipsehyperbola}) becomes$\colon$
 \begin{equation}\left\{\begin{array}{lr}x=c\cos \zeta \cosh u & ~\\
  y=c\sin \zeta \sinh u & ~\\
  \end{array}\right.~\label{hyperbola3}\end{equation}

 \noindent which is that of a hyperbola since  by using $\cosh^2 u -\sinh^2 u=1$, and $\cos^2 v+\sin^2 v=1$, we obtain equations~(\ref{hyperbola1}) and~(\ref{hyperbola2}) corresponding to a hyperbola and its characteristic relation respectively.\\
To prove that an ellipse and its confocal hyperbola are orthogonal, we define a functional $\phi(x,y)$ that emerges from equation~(\ref{ellipse1}) to be$\colon$
\begin{equation}\phi(x,y)=\displaystyle\frac{x^2}{a^2}+\displaystyle\frac{y^2}{b^2}-1=0~\label{ellipse4}\end{equation}
where $a$ is a constant equal to $c\cosh \eta$ and $b$ is another constant equal to $c\sinh \eta$.\\
  We define as well another functional $\psi(x,y)$ that stems out  from equation~(\ref{hyperbola1})  to be$\colon$
\begin{equation}\psi(x,y)=\displaystyle\frac{x^2}{a^2}-\displaystyle\frac{y^2}{b^2}+1=0~\label{hyperbola4}\end{equation}
where $a$ is a constant equal to $c\cos \zeta$ and $b$ is another constant equal to $c\sin \zeta$\,.\\

\noindent Now, $\vv\nabla\phi$ is a vector normal to the ellipse~\cite{piskonov}
\begin{equation}\vv{\nabla}\phi=2\displaystyle\frac{x}{a^2}\vv\imath+2\displaystyle\frac{y}{b^2}\vv\jmath=\displaystyle\frac{2c\cosh \eta\cos v}{c^2\cosh^2 \eta}\vv\imath
+\displaystyle\frac{2c \sinh \eta\sin  v}{c^2\sinh^2 \eta}\vv\jmath\end{equation}
\noindent Likewise, $\vv\nabla\psi$ is another vector   normal to the hyperbola
\begin{equation}\vv{\nabla}\psi=2\displaystyle\frac{x}{a^2}\vv\imath-2\displaystyle\frac{y}{b^2}\vv\jmath=\displaystyle\frac{2c\cos \zeta\cosh  u}{c^2\cos^2 \zeta}\vv\imath -\displaystyle\frac{2c\sin \zeta\sinh u}{c^2\sin^2 \zeta }\vv\jmath\end{equation}
Further, if the variable  $u$ takes the numerical value  $\eta$ and the variable  $v$ takes the numerical value $\zeta$, then from equations~(\ref{ellipse3}) and~(\ref{hyperbola3}) we have $x_{\mathrm{ellipse}}=x_{\mathrm{hyperbola}}$ and $y_{\mathrm{ellipse}}=y_{\mathrm{hyperbola}}$ and this is where the ellipse and the hyperbola intersect and thus it is evident that$\colon$ \begin{equation}\vv\nabla\phi\cdot\vv\nabla\psi=0 \label{phi}\end{equation}
Equation~(\ref{phi}) implies that the normals to the two curves are perpendicular to each other. Therefrom, and
on the basis of the fact that a normal to a curve at a given point is perpendicular to the tangent at that point, one can thus deduce that the tangents to the two curves are orthogonal
and hence the proof is complete.

\subsection*{Interference in a ripple tank} A ripple tank relates to a shallow tank of water conceived for the study of the properties of waves, such as, reflection, refraction, diffraction and interference. When a vibrator with one ball-ended dipper disturbs vertically the surface of water, a set of circular ripples are sent out in the medium. If the medium (water) is supposed devoid of viscosity, each molecule will vibrate in its place upon receiving the energy from the neighboring molecule lying on the same wave ray just behind it. Thus, a wave ray is a mathematical line that indicates the direction of propagation of the energy transported by the wave. However, the circular ripples observed on the surface of water and to which the wave rays are normal are called wavefronts. Correlatively each wavefront constitutes a set of points that are vibrating in phase and having the same number of wavelengths from the source.

Now, let's assume that the vibratory  state of the source is sinusoidal and of time equation $z=z_m\sin \omega t$. Hence, a water molecule $M$ on the surface lying at a distance $r$ from the source will receive the disturbance at a later time \, $t'=t+\displaystyle\frac{r}{v}$\,,  where $v$ is the velocity of propagation of the wave in the medium, and $\displaystyle\frac{r}{v}$ is specifically the time taken by the wave to reach $M$\,. As such, the vibratory state of $M$ at an instant $t$ will be similar to that of the source at an earlier stage and should be described by the relation$\colon$

\begin{equation}z=z_m\sin\omega\left(t-\displaystyle\frac{r}{v}\right) ~\label{vibratorystate}\end{equation}
  $z$ being the height of $M$ at the instant $t$ from the natural calm surface of water, $z_m$ is the amplitude of vibrations, and  $\omega$ is the pulsation of the vibration. Knowing that, $\omega$ is related to the frequency $f$ of vibrations by the relation $\omega=2\pi f$\,,  and that the wavelength is defined by $\lambda=\displaystyle\frac{v}{f}$\,, then, equation~(\ref{vibratorystate}) reduces to the form$\colon$
\begin{equation}z=z_m\sin\left(\omega t -\displaystyle\frac{2\pi r}{\lambda}\right)~\label{lambda}\end{equation}
Let's now consider the case where  $M$ receives two sinusoidal disturbances emanating from two sources $F_1$ and $F_2$ that are synchronous and coherent, and of separation $F_1F_2=2c$\,. The vibratory state of $M$ would then be determined by the principle of superposition~\cite{Hecht}, that suggests that the resultant disturbance at the point  $M$ is the algebraic sum of the separate constituent waves. i.e.
\begin{equation}
z=z_m\sin\left(\omega t -\displaystyle\frac{2\pi r_1}{v}\right) +z_m \sin \left(\omega t -\displaystyle\frac{2\pi r_2}{v}\right) ~\label{superposition1}
\end{equation}
Therefore
\begin{equation}
z_M=2z_m\cdot\cos\left(\displaystyle\frac{\pi(r_2-r_1)}{\lambda}\right)\cdot\sin\left(\omega t -\displaystyle\frac{\pi(r_2+r_1)}{\lambda}\right)~\label{superposition2}\end{equation}
 Let $A$ be    \begin{equation}A=2z_m\cos\left(\displaystyle\frac{\pi(r_2-r_1)}{\lambda}\right)\label{amplitude}\end{equation} In other words, $A$ is the part of equation~(\ref{superposition2}) that is independent of time  . Hence $A$  represents the amplitude of vibrations of the point $M$\,. Certainly the expression $\displaystyle\frac{\pi(r_2-r_1)}{\lambda}$ is a multiple of $\pi$, and can be written as
 \begin{equation}\displaystyle\frac{\pi(r_2-r_1)}{\lambda}=\alpha\pi \label{lambdapi}\end{equation}
 where $\alpha$ is a real algebraic number, and  by canceling $\pi$ from each side of equation~(\ref{lambdapi}) we obtain
 \begin{equation}r_2-r_1=\alpha\lambda ~\label{familyhyperbola} \end{equation}
 It goes without saying that equation~(\ref{familyhyperbola}) constitutes a continuous family of hyperbolae, each corresponding to a different value of $\alpha$\,.
 Furthermore, in observing the geometry of  Fig.1. one can easily notice  the triangle relation $|r_2-r_1|\leq F_1F_2$ which implies that
 \begin{equation}-F_1F_2\leq  r_2-r_1 \leq F_1F_2 \label{focus}\end{equation}
 and by substituting the value of $r_2-r_1$ given in equation~(\ref{familyhyperbola}) in equation~(\ref{focus}) we obtain
 \begin{equation}-\displaystyle\frac{F_1F_2}{\lambda}\leq \alpha \leq \displaystyle\frac{F_1F_2}{\lambda}~\label{kth}\end{equation}
 Consequently,
 \begin{equation}-\displaystyle\frac{2c}{\lambda}\leq \alpha \leq \displaystyle\frac{2c}{\lambda}~\label{kthh}\end{equation}
    Now we seek to find the parametric equations of the hyperbolic lines. For that purpose we compare equation~(\ref{literal}) with equation~(\ref{familyhyperbola}) and we obtain the expression of ($a$)  to be
      \begin{equation}a=\displaystyle\frac{\alpha\lambda}{2}~\label{alphalambda}\end{equation}
      The expression of ($a$) given by equation~(\ref{alphalambda}) should replace $c\cos \zeta$ in
  equation~(\ref{hyperbola3})\,, and by using relation~(\ref{hyperbola2}) we get  the expression of $b=\sqrt{\displaystyle\frac{{F_1F_2}^2}{4}-\displaystyle\frac{\alpha^2\lambda^2}{4}}$ that should replace the expression $c\sin \zeta$  in equation~(\ref{hyperbola3}) as well. And thus we obtain the parametric equations of the family of hyperbolae

  \begin{equation}\left\{ \begin{array}{lr}x=\displaystyle\frac{\alpha\lambda}{2}\,\cosh u & ~\\
 y=\, \sqrt{\displaystyle\frac{{F_1F_2}^2}{4}-\displaystyle\frac{\alpha^2\lambda^2}{4}}\,\sinh u & ~\\
  \end{array}\right. \label{finalhyperbolae}\end{equation}
  For each value of $\alpha$ there exists a hyperbolic line joining all the points that vibrate with the same amplitude at different phases.
    Now, if the real number $\alpha$ takes the value of an algebraic integer $k$, then by counting on relation~(\ref{amplitude}), the amplitude of the resultant wave at the point $M$ will take the highest possible value  $A=2y_m$\,, and correlatively the corresponding hyperbolae  are antinodal lines of parametric equations.

\begin{equation}\left\{ \begin{array}{lr}x=\displaystyle\frac{k\lambda}{2}\,\cosh u & ~\\
 y=\, \sqrt{\displaystyle\frac{{F_1F_2}^2}{4}-\displaystyle\frac{k^2\lambda^2}{4}}\,\sinh u & ~\\
  \end{array}\right. \label{antinodal}\end{equation}
   The number of antinodal lines would then be determined by the number of algebraic integers satisfying  equation~(\ref{kth})\,.\\

   If it happens that the real number $\alpha$ in equation ~(\ref{finalhyperbolae}) takes the value $\displaystyle\frac{2k+1}{2}$\,, where $k$ is an algebraic integer, then by counting again on relation~(\ref{amplitude}), the amplitude of the resultant wave at the point $M$ will be zero, hence the point $M$ will not vibrate at all. Correspondingly, the observed hyperbolic lines are  nodal and of parametric equations

   \begin{equation}\left\{ \begin{array}{lr}x=\displaystyle\frac{(2k+1)\lambda}{4}\,\cosh u & ~\\
 y=\, \sqrt{\displaystyle\frac{{F_1F_2}^2}{4}-\displaystyle\frac{(2k+1)^2\lambda^2}{16}}\,\sinh u & ~\\
  \end{array}\right. \label{finalhyperbolas}\end{equation}

    \noindent The number of nodal lines would then be determined by replacing $\alpha$ in inequality~(\ref{kth}) by the expression $\displaystyle\frac{2k+1}{2}$\, to obtain    \begin{equation}-\displaystyle\frac{F_1F_2}{\lambda}-\displaystyle\frac{1}{2}\leq k\leq \displaystyle\frac{F_1F_2}{\lambda}-\displaystyle\frac{1}{2}\label{minima}\end{equation}
Hence the number of nodal lines would then be determined by the number of algebraic integers satisfying inequality~(\ref{minima}).\\

We have thus far established the parametric equations of the hyperbolic lines, each of which representing the set of points vibrating with the same amplitude at different phases. Our intent in this section is to develop the parametric equations of the lines, that individually  represents the set of points vibrating in phase but at different amplitudes. For that purpose we consider the expression of the initial phase $\displaystyle\frac{\pi(r_2+r_1)}{\lambda}$ given in equation~(\ref{superposition2}), clearly this  initial phase is a multiple of $\pi$\,, and it can be written as
\begin{equation} \displaystyle\frac{\pi(r_2+r_1)}{\lambda}=\beta \pi \label{multiple} \end{equation}
where $\beta$ is a real  number, and by canceling $\beta$ from equation~(\ref{multiple}) we obtain
\begin{equation} r_2+r_1=\beta \lambda \label{multiple1} \end{equation}
Evidently, equation~(\ref{multiple1}) is that of a continuous family of ellipses each corresponding to a different value of $\beta$\,. Furthermore, in observing the geometry of Fig.1, one can clearly notice the triangle relation $r_2+r_1\geq F_1F_2$ which implies that $\beta \lambda \geq 2c$ and finally $\beta\geq\displaystyle\frac{2c}{\lambda}$\,.\\
Now we seek to find the parametric equations of the elliptic lines, for that purpose we compare equation~(\ref{ellipse1}) with equation~(\ref{multiple1}) and we get the expression of ($a$) to be
\begin{equation}a=\displaystyle\frac{\beta \lambda}{2} \label{betalambda} \end{equation}
The expression of the term ($a$) given by equation~(\ref{betalambda}) should replace that of $c\cosh \eta$ in equation~(\ref{ellipse3}) and by using equation~(\ref{ellipse2}) we get the expression of $b=\sqrt{\displaystyle\frac{\beta^2 \lambda^2}{4}-\displaystyle\frac{{F_1F_2}^2}{4}}$ that should replace $c\sinh \eta$ in equation~(\ref{ellipse3}), hence we obtain the parametric equations of the family of ellipses

   \begin{equation}\left\{ \begin{array}{lr}x=\displaystyle\frac{\beta\lambda}{2}\,\cos v & ~\\
 y=\, \sqrt{\displaystyle\frac{\beta^2\lambda^2}{4}-\displaystyle\frac{{F_1F_2}^2}{4}}\,\sin v & ~\\
  \end{array}\right. \label{finalellipses}\end{equation}
\section*{Conclusion}
 In reality, equation~(\ref{finalellipses}) is that of a family of ellipses, each taken separately is a wave front comprising two types of alternating  bundles of points vibrating in opposite phases, owing to the change in the sign of the amplitude caused by its  position dependence. Every individual bundle of points that are vibrating in phase is orthogonal to a set of hyperbolae representing points vibrating with the same amplitude at different phases.  As such and on the bases of the fact that wave rays are orthogonal trajectories of the wavefronts~\cite{Hecht}, one can conclude that the hyperbolic lines defined by equation~(\ref{finalhyperbolae}) are nothing but curved  wave rays of the resultant wave.\\
\mbox{\includegraphics[]{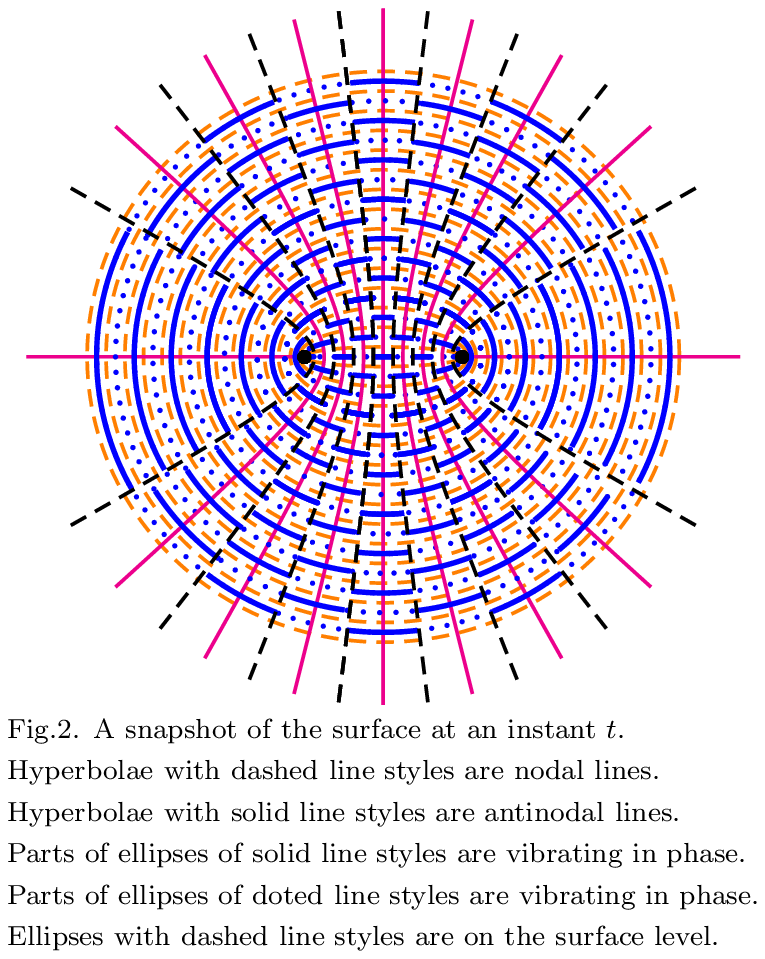}\hfil\includegraphics[]{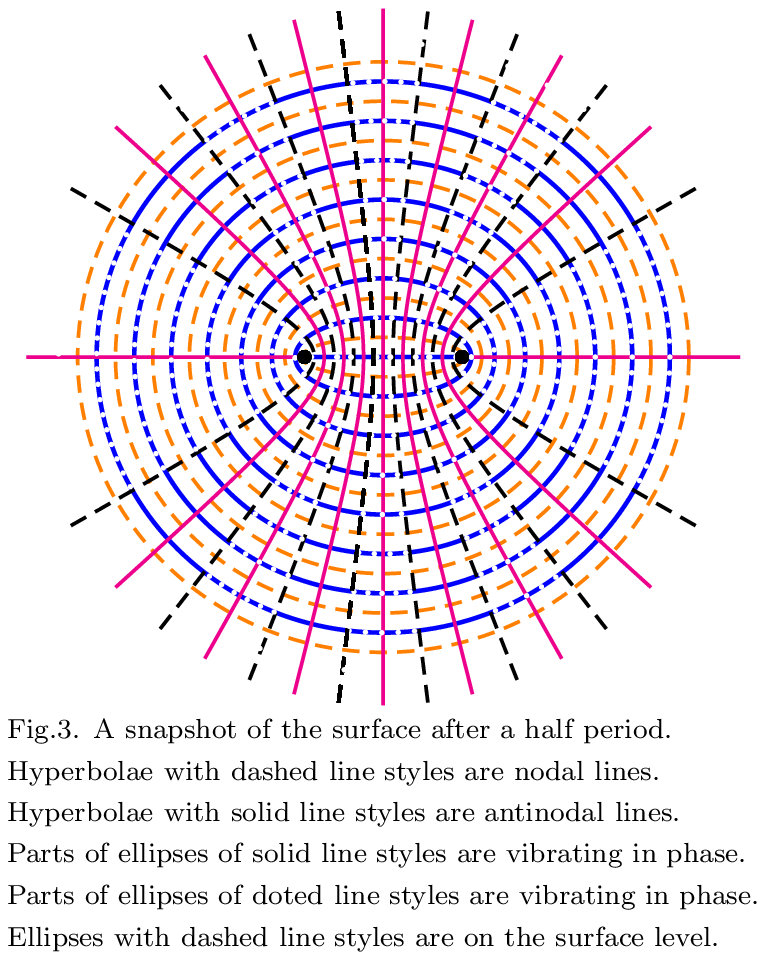}}

\begin{center}
\includegraphics[]{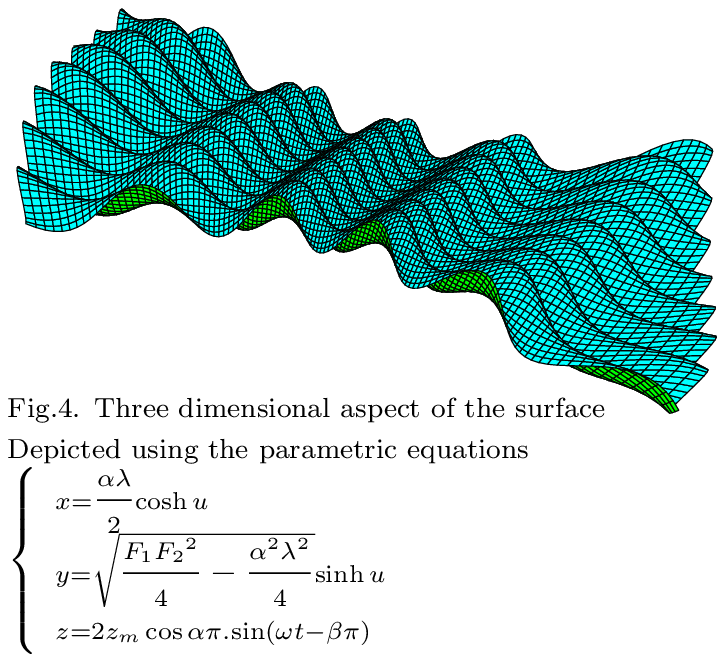}
\end{center}

\end{document}